\documentclass{article}

\usepackage{PRIMEarxiv}
\usepackage{listings}
\usepackage{courier}

\usepackage[utf8]{inputenc} 
\usepackage[T1]{fontenc}    
\usepackage{hyperref}       
\usepackage{url}            
\usepackage{booktabs}       
\usepackage{amsfonts}       
\usepackage{nicefrac}       
\usepackage{microtype}      
\usepackage{lipsum}
\usepackage{url}
\usepackage{fancyhdr}       
\usepackage{graphicx}       
\graphicspath{{media/}}     

\title{Exploring Durham University Physics exams with Large Language Models}

\author{
  Will Yeadon \\
  Department of Physics \\ 
  Durham University \\
  United Kingdom \\
  \texttt{will.yeadon@durham.ac.uk} \\
  \And
  Douglas P. Halliday \\
  Department of Physics \\ 
  Durham University \\
  United Kingdom \\
  \texttt{d.p.halliday@durham.ac.uk} \\
}

\begin{document}
\maketitle

\begin{abstract}
The emergence of advanced Natural Language Processing (NLP) models like ChatGPT has raised concerns among universities regarding AI-driven exam completion. This paper provides a comprehensive evaluation of the proficiency of GPT-4 and GPT-3.5 in answering a set of 42 exam papers derived from 10 distinct physics courses, administered at Durham University over the span of 2018 to 2022, totalling 593 questions and 2504 available marks. These exams, spanning both undergraduate and postgraduate levels, include traditional pre-COVID and adaptive COVID-era formats. Questions from the years 2018-2020 were designed for pre-COVID in person adjudicated examinations whereas the 2021-2022 exams were set for varying COVID-adapted conditions including open-book conditions. To ensure a fair evaluation of AI performances, the exams completed by AI were assessed by the original exam markers. However, due to staffing constraints, only the aforementioned 593 out of the total 1280 questions were marked. GPT-4 and GPT-3.5 scored an average of 49.4\% and 38.6\%, respectively, suggesting only the weaker students would potential improve their marks if using AI. For exams from the pre-COVID era, the average scores for GPT-4 and GPT-3.5 were 50.8\% and 41.6\%, respectively. However, post-COVID, these dropped to 47.5\% and 33.6\%. Thus contrary to expectations, the change to less fact-based questions in the COVID era did not significantly impact AI performance for the state-of-the-art models such as GPT-4. These findings suggest that while current AI models struggle with university-level Physics questions, an improving trend is observable. The code used for automated AI completion is made publicly available for further research.
\end{abstract}

\keywords{ChatGPT \and AI \and GPT-4 \and COVID}

\section{Introduction}
The rise of Natural Language Processing (NLP) models like ChatGPT has prompted concerns about the potential for both coursework assignments and exams to be automatically completed with a high degree of accuracy by AI models. As these models continue to advance in their ability to process and analyze natural language, it's becoming increasingly important to evaluate their efficacy in answering complex questions, particularly in academic fields such as Physics. Freely available software, such as ChatGPT, is capable of producing high-quality written output that is both coherent and accurate. The technical report for GPT-4 (the model driving the paid version of ChatGPT) has demonstrated that it outperforms the vast majority of human participants on various professional and academic exams, including the Scholastic Assessment Test (SAT), Graduate Record Examination (GRE), Advanced Sommelier exam, and Advanced Placement (AP) courses such as US Government, Psychology, and Environmental Science \cite{gpt4TechnicalReport}. Furthermore, recent independent publications have confirmed that natural language processing (NLP) models, like ChatGPT, are able to pass professional exams in the medical and legal fields \cite{medicalPerformance, lawPerformance}.

ChatGPT has demonstrated impressive capabilities in the domain of Physics, even scoring the highest possible grade on a current UK university Physics module in an essay-based assessment \cite{Yeadon_2023}. However, separate studies have shown it struggling to answer questions that necessitate an appreciation of the underlying Physics \cite{Socrates, gptPhysicsCourse}. With the advent of the API for GPT-3.5 and GPT-4, it is now possible to automate the process of querying these models with a large volume of questions, without the need for a one-by-one, chat-style interaction. Leveraging this capability, the current paper aims to comprehensively evaluate the performance of these state-of-the-art large language models (LLMs) in solving real-world, university-level Physics questions that have appeared in actual exams at Durham University thus quantifying the potential risks to assessment integrity. 

The the efficacy of large language models (LLMs) varies depending on the prompting techniques. This variance implies that the performance can be enhanced or hindered if the prompt is strategically constructed knowing the desired answer or manipulated to sabotage. To simulate a real-world scenario where a student might use AI assistance during exams, questions were posed in a zero-shot manner along with system messages using OpenAI's API. This is reflective of a good prompt \cite{openai_best, fewShot} whilst remaining within the API token limit. Further, by having the AI-generated responses assessed by the same academics who evaluate student submissions, this research provides a robust evaluation of AI performance in an authentic university context. For full transparency and replicability, all code used in this research is freely available on GitHub\footnote{https://github.com/WillYeadon/AI-Exam-Completion}.

\subsection{Related work}
\label{related-work}
Previous assessments of the Physics capabilities of advanced AI models have often focused narrowly on specific problems or individual courses \cite{physPrev1, physPrev2}. There has also been research exploring the mathematical capabilities of these models on a larger scale \cite{mathsCapChat1, mathsCapChat2}. Nevertheless, many university-level Physics questions demand an understanding of the real-world phenomena under investigation, not merely mathematical manipulation. This requirement sets the present study apart from prior work, which has predominantly focused on mathematical problem-solving skills.

\section{Methodology}
\label{sec:method}
\subsection{Overview}
To automate the completion of exams with AI, we utilized a series of regular expressions for identifying and extracting questions within the '.tex' files used to create Physics exams at Durham University. Each file corresponds to a single numerical question. Typically, the exams are comprised of 4 - 6 such numerical questions, each containing 4 - 8 alphabetically organized sub-questions. Within each alphabetical sub-question, further roman-numeral itemized questions may be present, although there are instances of non-itemized sub-questions within an alphabetical sub-question. In addition, some questions feature a preamble providing relevant information preceding the main question, or include graphics related to the question.

Given this wide variety of question formats, the regular expressions do not consistently capture each question perfectly, occasionally registering irrelevant surrounding text as part of the question. In addition to the question itself, the question number, the presence or absence of graphics within the question, and the name of the '.tex' file containing the question are all recorded. Upon extraction, an initial cleaning process is performed on the questions. This process involves the deletion of commonly captured LaTeX fragments such as \emph{documentclass}, \emph{setlength}, and \emph{hsize}, which are often incorrectly registered as part of the questions. The modified questions are then sent to OpenAI's API, where GPT-3.5 is instructed to amend the LaTeX if it contains any errors that would prevent compilation.

Concurrent with these steps, additional scripts are employed to extract the marks allocated to each question and verify the correctness of the recorded question numbers. It is important to note that these steps are followed by a manual checking procedure, during which each extracted question is inspected to ensure the scripts functioned as expected. For this study, approximately $5\%$ of the questions required manual modification. Subsequently, a second script sends the cleaned questions, along with system prompts, back to the API for the generation of answers. This process includes an additional LaTeX compilation check. Further details of the AI answering process can be found in Section \ref{sec-api}. The questions, answers, and marks are then compiled to create a unique PDF for each exam. These exams were subsequently marked by the academics responsible for each module.

\subsection{Dataset}
Between 2018 and 2022, a total of 84 Physics exams were administered at Durham University, yielding approximately 2410 individual questions that were successfully extracted and sent to OpenAI's API. However, due to staffing constraints, only 42 out of these 84 exams had at least one section marked (or 593 out of 2410 questions). Nevertheless, care was taken to ensure that at least one exam was assessed from each level of the physics degree program, encompassing the three years of a Bachelor's degree (Levels 1, 2, and 3) and the additional year for a Master's level (Level 4). This distribution is summarized in Table \ref{exam-list}.

The original exams and their respective model solutions can only be accessed through a Durham University staff or student account, which significantly reduces the likelihood of these questions having been memorized by the AI models during training. Yet, considering the vast availability of academic Physics content online, it's plausible that questions of similar nature were part of the training data for GPT-3.5 or GPT-4.

Regardless of any potential data overlap, the primary objective of this study is to evaluate the proficiency of these AI models in generating answers to Physics questions. The possibility that the AI models might have previously encountered similar questions during training doesn't substantially undermine this objective. In fact, the effectiveness of AI in academic contexts is often determined by its ability to apply learned patterns to analogous but distinct problems. Hence, even if some questions were part of the models' training data, their successful answering of these questions serves as evidence of their utility in educational environments.

\begin{table}[ht]
  \centering
  \caption{Full list of exams marked for the present study.}
  \label{exam-list}
  \begin{tabular}{@{}lccccc@{}}
    \toprule
    \textbf{Exam Paper}                   & \textbf{Level} & \textbf{Years} & \textbf{Models} & \textbf{Number of} & \textbf{Available} \\ 
    										  & 	            &                & \textbf{Used} & \textbf{Questions} & \textbf{Marks} \\
    \midrule
    Foundations of Physics 1              & 1              & 2020 - 2022              & GPT-4, GPT-3.5 & 22                  & 75                        \\
    Foundations of Physics 2A             & 2              & 2018 - 2022              & GPT-4, GPT-3.5 & 65                  & 306                        \\
    Theoretical Physics 2                 & 2              & 2018 - 2022              & GPT-4, GPT-3.5 & 80                  & 302                        \\
    Mathematical Methods in Physics       & 2              & 2018 - 2022              & GPT-4, GPT-3.5 & 80                  & 271                        \\
    Foundations of Physics 3A             & 3              & 2018 - 2022              & GPT-4, GPT-3.5 & 69                  & 222                        \\
    Planets and Cosmology                 & 3              & 2019 - 2022              & GPT-4, GPT-3.5 & 49                  & 220                        \\
    Modern Atomic and Optical Physics 3   & 3              & 2019 - 2022              & GPT-4          & 37                  & 186                        \\
    Theoretical Physics 3                 & 3              & 2019 - 2022              & GPT-4          & 59                  & 248                        \\
    Advanced Condensed Matter Physics     & 4              & 2018 - 2022              & GPT-4, GPT-3.5 & 111                 & 554                        \\
    Theoretical Astrophysics              & 4              & 2021 - 2022              & GPT-4          & 21                  & 120                        \\ 
    \midrule
    Mean                                  &                &                          &                & 59.3                & 250.4                      \\
    Total                                 &                &                          &                & 593                 & 2504                       \\
    \bottomrule
  \end{tabular}
\end{table}

\subsection{AI answers}
\label{sec-api}
The AI models answered the questions through the use of OpenAI's API. This process adhered to commonly accepted best practices and guidelines \cite{openai_best, fewShot}, and each question included a system message instructing the AI model to assume the role of a Physics professor and to use LaTeX for generating answers. The complete system message is illustrated in Figure \ref{system-message-answer}.

Once the AI model provided a response, it was compiled into a temporary PDF through Jinja \cite{jinja}. If the response failed to compile, it was returned to the AI model along with a request to correct the LaTeX code. This process could iterate up to three times before substituting the response with a 'Compilation failed' message. For both AI models, at least one retry was necessary for approximately 12.02\% of the extracted questions. However, 'Compilation failed' messages occurred more frequently with GPT-3.5 (3.04\% of responses) compared to GPT-4 (0.33\% of responses).

Many of the sub-questions within the selected exams referred to earlier sub-questions (e.g., 'using your answer to part a)...'). In order to provide accurate answers, the AI models required access to these previous questions and their corresponding responses. This need was initially addressed by appending prior question-answer pairs to the API requests using a sliding-window technique. However, given the retry rate of 12.02\%, this method quickly resulted in the total length of messages surpassing the per-request limit set by the API. Therefore, a new approach was adopted in which each question was treated independently. The exam markers were then instructed to identify and record instances of 'question error,' which occurred when the AI models lacked access to necessary information due to this isolation. This procedure also applied when the questions included graphics.

\begin{figure}[ht]
    \centering
    \begin{lstlisting}[breaklines=true, basicstyle=\footnotesize\ttfamily, frame=single]
    You are a Physics professor.
    Answer the question between triple backticks, showing your work step by step.
    IMPORTANT: Use Latex code for equations and numbers. Please define any custom LaTeX commands you use in your answer.
    IMPORTANT: Ensure numbers and units in the text are properly formatted using LaTeX, e.g., use `$10^24 cm^{-2}$` instead of '10^24 cm^-2'.
    IMPORTANT: Do not attempt to draw sketches or images. Do not include figures.
    IMPORTANT: Limit your answers to a maximum of 6000 characters (approx. 1500 tokens).
    \end{lstlisting}
    \caption{The full system message used when sending the exam questions to the API.}
    \label{system-message-answer}
\end{figure}


\section{Results}
\label{results}
\begin{figure}[!htp]
\centering
\includegraphics[width=13.5cm]{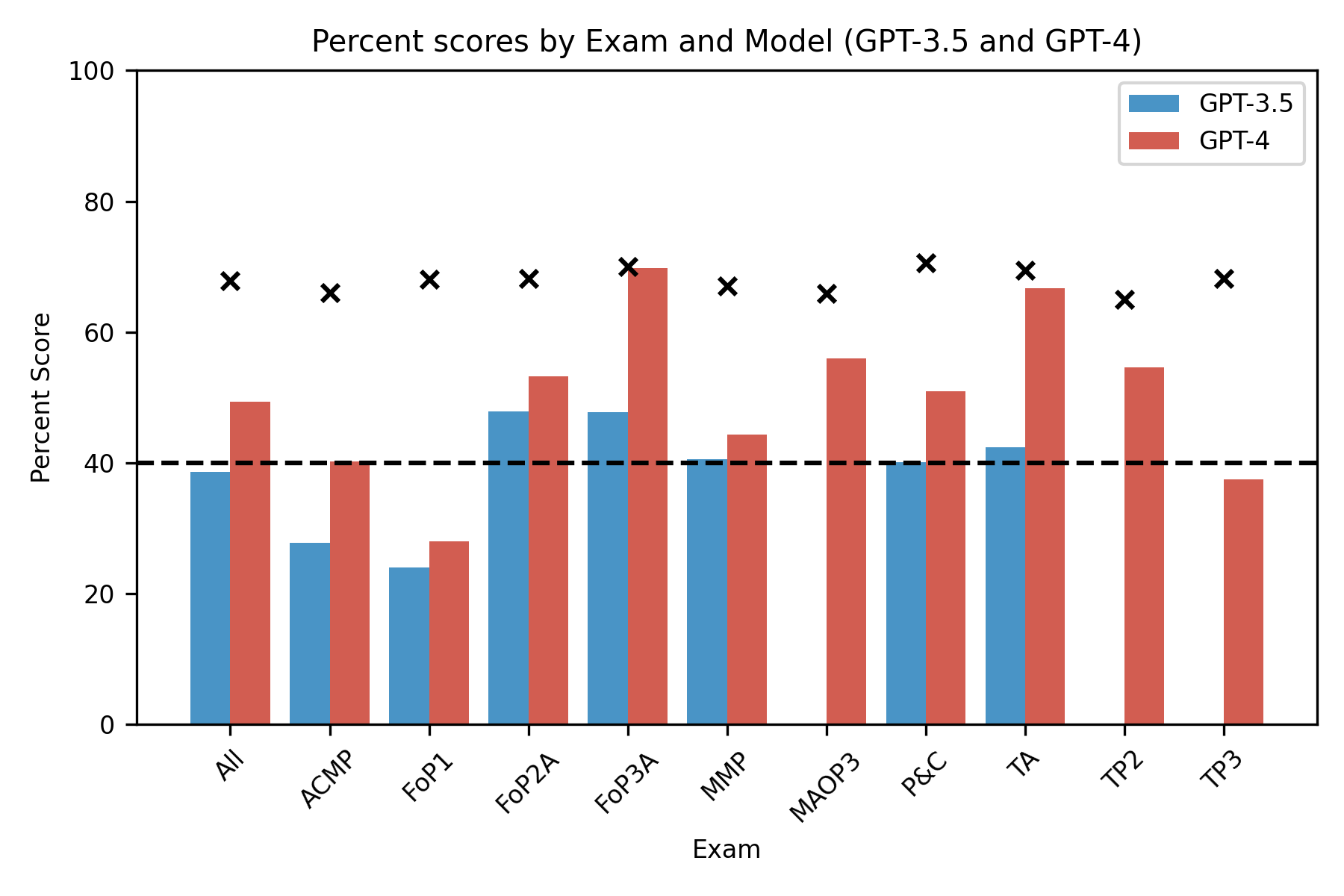}
\caption{Performance of GPT-4 and GPT-3.5 on different physics exams. The black crosses indicate the average student mark from 2018 - 2021 on the modules for the exam and the dashed black line shows the 40\% score required to pass the exam.}
\label{fig-all}
\end{figure}

The performance of both GPT-4 and GPT-3.5 on a diverse set of Physics examinations was evaluated. A summary of the AI model scores on each exam is illustrated in Figure~\ref{fig-all}. The exams are denoted by abbreviations, each corresponding to a unique course or exam set. The explanation for each abbreviation is provided in Table~\ref{abbrv-table}. As evident from Figure~\ref{fig-all}, GPT-4 consistently outperformed GPT-3.5 across all exam types, with an average score of 49.4\% compared to 38.6\% for GPT-3.5. For the vast majority of exams GPT-4 scores considerably lower than the average student scores from 2018 - 2021 \cite{foi_2018, foi_2019, foi_2020, foi_2021} however it does get close for Foundations of Physics 3A and Theoretical Astrophysics.    

\begin{table}[ht]
\centering
\caption{Abbreviations for Exam Names}
\label{abbrv-table}
\begin{tabular}{ll}
\toprule
\textbf{Abbreviation} & \textbf{Full Exam Name} \\
\midrule
All & All exam results combined \\
FoP1 & Foundations of Physics 1 \\
FoP2A & Foundations of Physics 2A \\
TP2 & Theoretical Physics 2 \\
MMP & Mathematical Methods in Physics \\
FoP3A & Foundations of Physics 3A \\
P\&C & Planets and Cosmology \\
MAOP3 & Modern Atomic and Optical Physics 3 \\
TP3 & Theoretical Physics 3 \\
ACMP & Advanced Condensed Matter Physics \\
TA & Theoretical Astrophysics \\
\bottomrule
\end{tabular}
\end{table}

\begin{figure}[!htp]
\centering
\includegraphics[width=13.5cm]{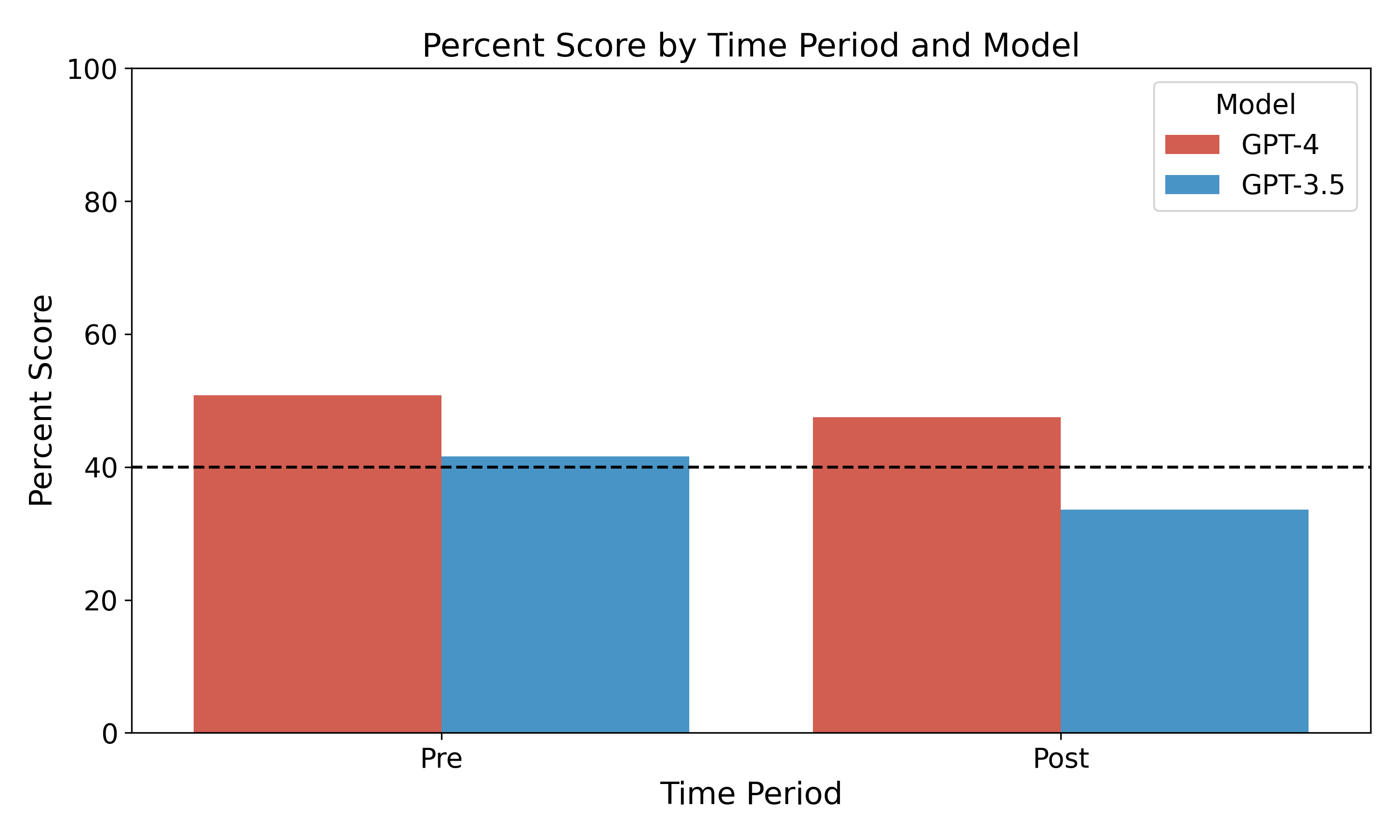}
\caption{Performance of GPT-3.5 and GPT-4 in pre and post-COVID exams. The dashed black line shows the 40\% score required to pass the exam.}
\label{fig-covid}
\end{figure}

Additionally, the performance of both AI models before and after the emergence of COVID-19 was investigated (Figure~\ref{fig-covid}). In response to the pandemic, many traditional in-person adjudicated exams transitioned to open-book conditions and the questions were adapted accordingly. On the pre-COVID exams, GPT-4 and GPT-3.5 scored 50.8\% and 41.6\% respectively whereas post-COVID the decreased to 47.5\% and 33.6\% respectively. Contrary to expectations that AI models would perform significantly worse with less 'factual recall' style questions, particularly for GPT-4, the decrease in scores in the post-COVID era was only slight.

\begin{figure}[!htp]
\centering
\includegraphics[width=13.5cm]{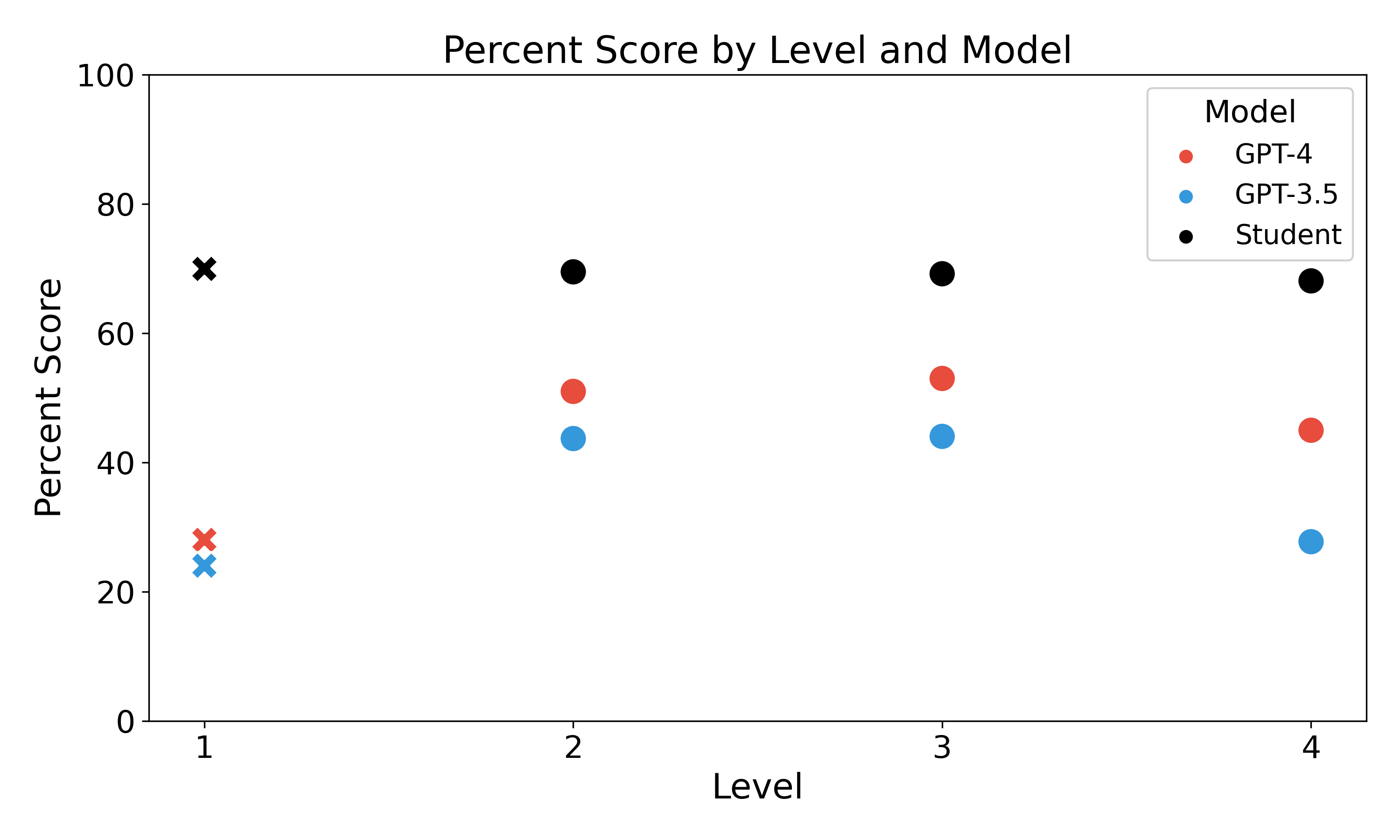}
\caption{Performance of GPT-3.5 and GPT-4 across different exam levels. As there was only 22 first year questions in the study, the level 1 markers are shown with a 'x' as they are unlikely to be fully representative. The Student series in black in calculated from the average student mark between 2018 and 2021.}
\label{fig-level}
\end{figure}

Lastly, an analysis of the AI models' performance across different exam levels was conducted (Figure~\ref{fig-level}). Here, the 'levels' represent the progression through university studies, from the first year of the Bachelor's program (level 1) through to the Master's degree (level 4). It is reasonable to expect performance to deteriorate as the complexity of the questions increases (in line with the increase in level) however there doesn't appear to be a clear trend. 

\section{Discussion}
\subsection{Risk to assessment fidelity}
The current study indicates that neither GPT-4 nor GPT-3.5 pose a significant threat to the fidelity of Physics assessments, as the performance of these AI models would only achieve an average 'pass' grade or below in most cases. This aligns with the broader context of undergraduate assessment at Durham University, where exams comprise only a portion of the overall grade. Other components of a Physics degree, such as complex coding problems and essay writing, could potentially present more significant challenges given prior work demonstrating AI proficiency in these areas \cite{Yeadon_2023}.

While it has been established that humans struggle to reliably detect AI-generated text \cite{humanDetection1, humanDetection2, humanDetection3}, certain detection technologies claim to counter this issue. However, skepticism is warranted, as many detection tools operate on the principle of using portions of the text to predict subsequent words and match them with the actual text. This approach is susceptible to 'paraphrasing attacks,' where reordering of words can bypass the detection mechanism \cite{detectAI}. Moreover, this 'next word' prediction style carries a risk of bias against non-native English speakers \cite{englishBias}.

Based on these findings, it seems that the immediate risk to assessment fidelity in Universities is relatively low. Nevertheless, there are certain assessment elements where the risk could be more substantial. Given the rapid advancement of AI technologies, it is crucial for universities to remain vigilant, routinely monitoring AI capabilities using the method outlined in this study or other equally robust approaches. 

\subsection{Factors Impacting AI Performance}
\label{sec-fail}
Common sense assumptions of AI models performing worst on more complex problems (Masters vs Bachelors) or worse on questions written for open-book do not seem to be born out in the data. To delve deeper into the factors that could potentially impact the performance of AI models, a word analysis was conducted, which focused on the fraction of marks achieved by the AI model GPT-4 - the best performing of the two models. The analysis evaluated the correlation between the appearance of certain words in the questions and the corresponding scores achieved by the AI, both for low (<40\%, a fail at UK Universities) and high scoring (>90\%) responses.

Though the correlations were not exceptionally strong, a discernible pattern did emerge. Terms with a mathematical connotation such as 'calculate', 'partial', and 'frac' were generally associated with lower scores. On the other hand, the term 'explain' and names often used in physics nomenclature, like '$m_1$' and '$m_2$', were linked to higher scores. This could potentially be indicative of the AI's stronger performance in conceptual understanding and explication rather than mathematical computation. However, again, the correlations are low so this is merely indicative.

Of note, the word 'sketch' was strongly negatively correlated with the AI's score. This was the strongest negative correlation among all the analyzed words. This is unsuprising given the API is currently text-based so it can only sketch by printing out lines of symbols. Another interesting finding was the limited impact of question length on the AI's performance. The correlation between the length of the question and the fraction of marks achieved was a mere 0.061, implying that the length of the question does not significantly influence the AI's ability to generate correct answers.

\begin{figure}[ht]
\centering
\includegraphics[width=0.45\textwidth]{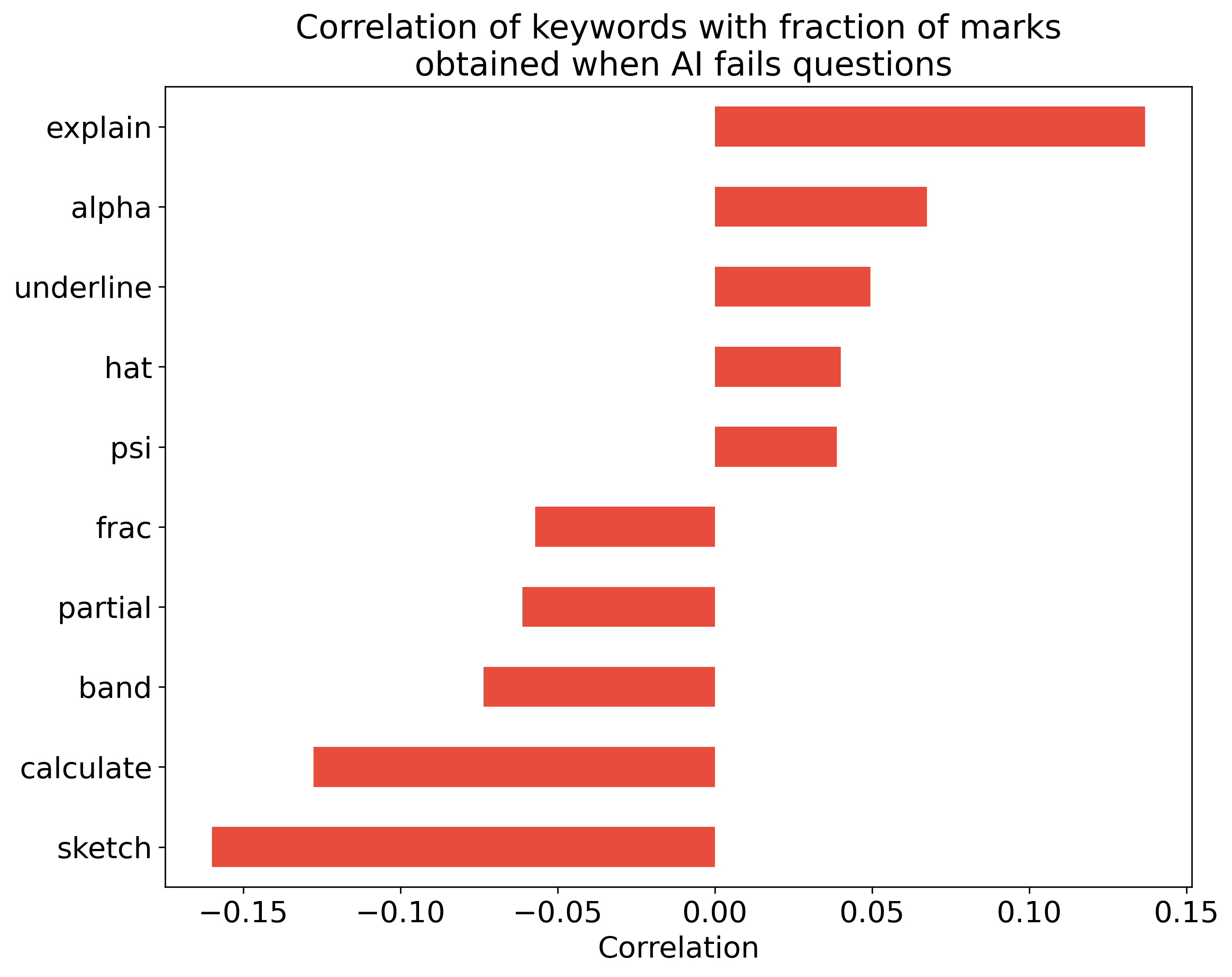}
\includegraphics[width=0.45\textwidth]{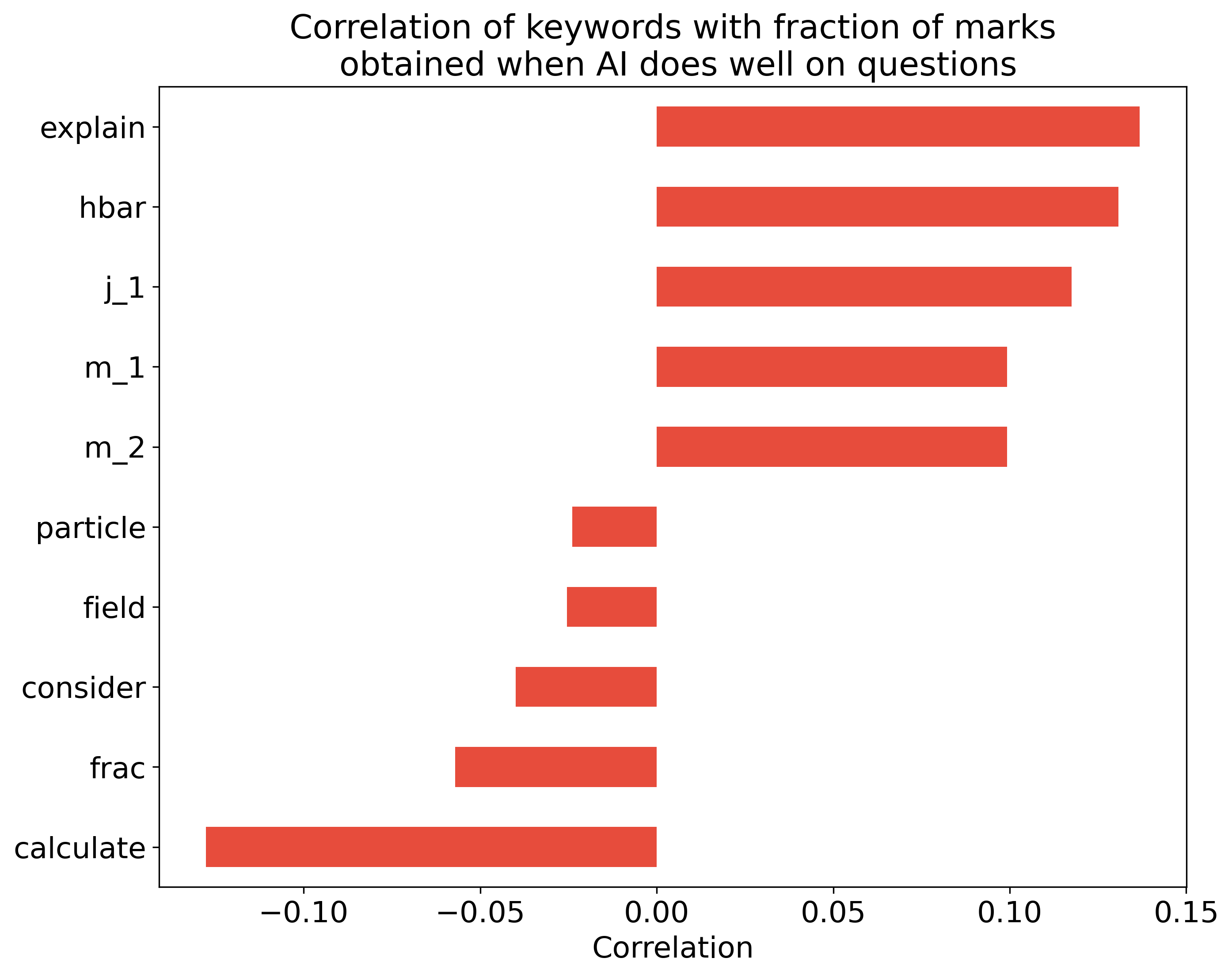}
\caption{Correlation of keywords with fraction of marks obtained when AI fails (left) and does well (right) on questions.}
\end{figure}

\subsection{Use cases}
\begin{figure}[!htp]
\centering
\includegraphics[width=13.5cm]{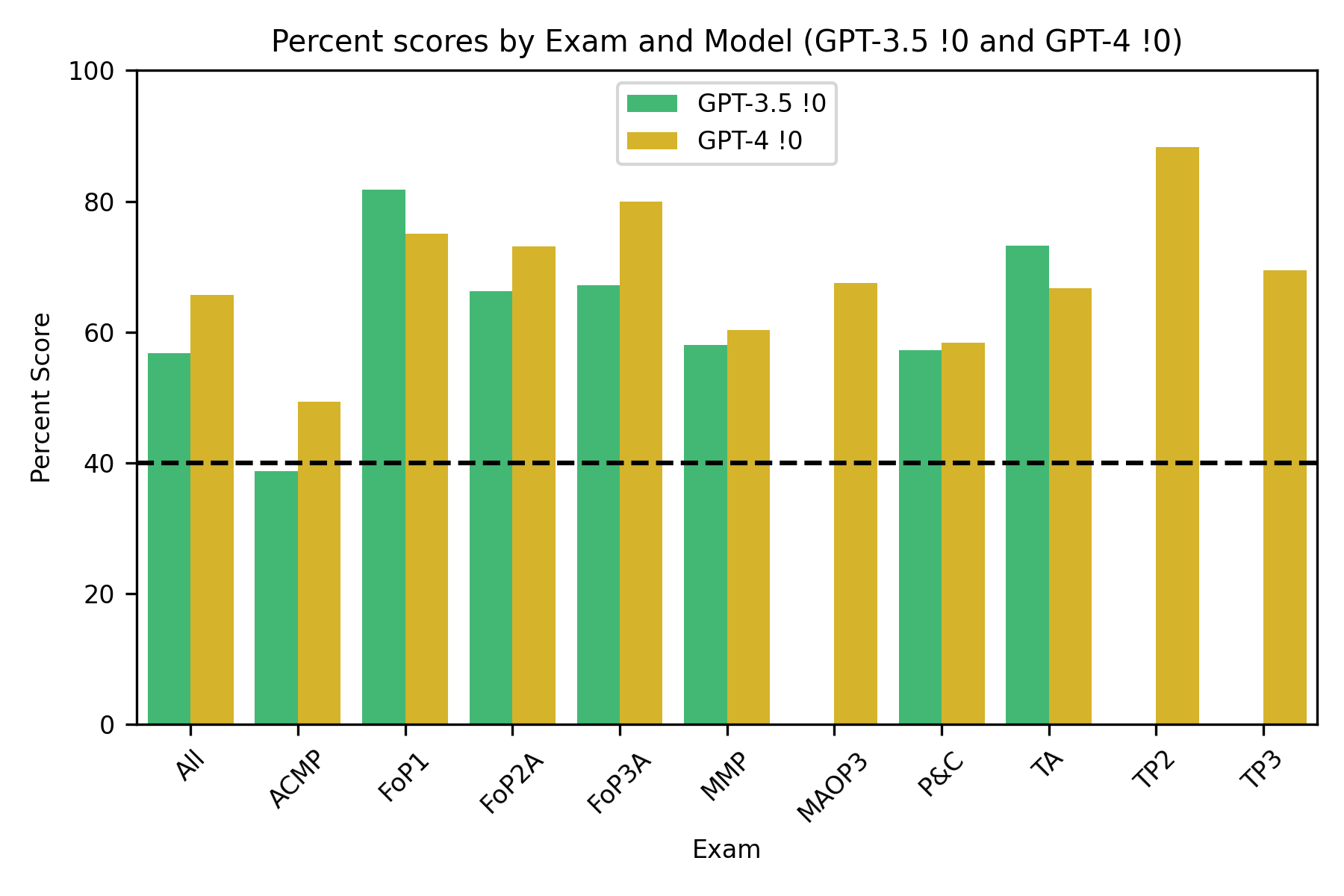}
\caption{Percent scores achieved by the AI models GPT-3.5 and GPT-4 when excluding answers which the models scored zero on.}
\label{fig-_0}
\end{figure}

Beyond the immediate threat to exam integrity, there is potential for AI to be integrated within Physics education as a tool to aid students. At first glance, the AI's low scores might seem to dismiss this idea. However, instances where the AI scored zero marks can be viewed as scenarios where the AI either misinterpreted the question or was unable to answer it. These instances occurred 30.0\% of the time for GPT-4 and 40.1\% of the time for GPT-3.5. If we exclude these zero-scoring answers, our analysis shifts to examine the AI models' performance in scenarios where they appear to understand the question. This perspective is illustrated in Figure \ref{fig-_0}. Unsurprisingly, these scores are significantly higher than those obtained when considering all attempts. The overall averages notably stand at 56.7\% for GPT-3.5 and 65.6\% for GPT-4, indicating a potential use case for AI as a provider of useful 'hints' that students could expand upon. Interestingly, when it came to Mathematical Methods in Physics, GPT-3.5 outperformed GPT-4, marking the only instance of such a result.

\section{Conclusion}
The study demonstrates that whilst there is potential for AI models to complete Physics exams, currently universities should be aware but should not panic. Whilst there are instances of strong performance, in general only the very weakest students would benefit from having AI complete all of their work. However, given the clear improvement from GPT-3.5 to GPT-4, it's not a stretch to imagine future models posing a substantial challenge to exam integrity. Further, it should be noted that these results were for Physics exams yet Physics degrees often have other have other components such as essay writing and coding assignments which are considerably more vulnerable to automatic AI completion. 

It is recommend educators should consider using the open-source code provided\footnote{https://github.com/WillYeadon/AI-Exam-Completion} to perform their own risk assessment for their exams and for transparency with students so that they are aware of whether to trust AI output. The slight drop in AI performance on post-COVID exams, attributable to adapted remote exam questions, is not enough to conclude more 'understanding-based' questions will stop AI completion especially given how GPT-4 post-COVID scored better than GPT-3.5 pre-COVID. In the short term, it is recommended that educators consider question styles that AI models currently struggle with: multipart problems, sketch-based questions, and queries requiring the interpretation of graphics. However, these recommendations are provisional, as AI performance is likely to continue improving rapidly. 

A longer term strategy may involve open-book exams requiring to be set such that a modern AI model would score a maximum of 40\% (the typical pass mark at UK universities) before being approved. Furthermore, implementing more invigilated, in-person exams could reinforce exam integrity. Beyond these immediate solutions, the predicament offers an opportunity for education to evolve its assessment methods. The focus could shift away from traditional exams and instead embrace a more holistic, skills-based approach, certifying that students have reached a 'sufficient level' of competency in various aspects of Physics.

\section*{Acknowledgments}
This work greatly benefited from the invaluable contributions of our academic colleagues who generously dedicated their time to assess the AI-generated responses. Our profound gratitude goes to Ian Terry, Fernando Dias, Simon Morris, Christine Done, Nikitas Gidopoulos, Jeppe Andersen, Robert Potvlige, David Carty, Craig Testrow, and Vincent Eke. Their expertise and meticulous attention to detail were instrumental in bringing this research to fruition.

\bibliographystyle{unsrt}  
\bibliography{references}

\begin{thebibliography}{10}

\bibitem{gpt4TechnicalReport}
OpenAI.
\newblock Gpt-4 technical report.
\newblock {\em ArXiv}, abs/2303.08774, 2023.

\bibitem{medicalPerformance}
Tiffany~H Kung, Morgan Cheatham, Arielle Medenilla, Czarina Sillos, Lorie
  De~Leon, Camille Elepa{\~n}o, Maria Madriaga, Rimel Aggabao, Giezel
  Diaz-Candido, James Maningo, et~al.
\newblock Performance of chatgpt on usmle: Potential for ai-assisted medical
  education using large language models.
\newblock {\em PLoS digital health}, 2(2):e0000198, 2023.

\bibitem{lawPerformance}
Jonathan~H Choi, Kristin~E Hickman, Amy Monahan, and Daniel Schwarcz.
\newblock Chatgpt goes to law school.
\newblock {\em Available at SSRN}, 2023.

\bibitem{Yeadon_2023}
Will Yeadon, Oto-Obong Inyang, Arin Mizouri, Alex Peach, and Craig~P Testrow.
\newblock The death of the short-form physics essay in the coming ai
  revolution.
\newblock {\em Physics Education}, 58(3):035027, April 2023.

\bibitem{Socrates}
Bor Gregorcic and Ann-Marie Pendrill.
\newblock Chatgpt and the frustrated socrates.
\newblock {\em Physics Education}, 58(3):035021, March 2023.

\bibitem{gptPhysicsCourse}
Gerd Kortemeyer.
\newblock Could an artificial-intelligence agent pass an introductory physics
  course?
\newblock {\em Physical Review Physics Education Research}, 19(1):010132, 2023.

\bibitem{openai_best}
OpenAI.
\newblock Best practices for prompt engineering with openai api, June 2023.
\newblock
  \url{https://help.openai.com/en/articles/6654000-best-practices-for-prompt-engineering-with-openai-api}.

\bibitem{fewShot}
Tom Brown, Benjamin Mann, Nick Ryder, Melanie Subbiah, Jared~D Kaplan, Prafulla
  Dhariwal, Arvind Neelakantan, Pranav Shyam, Girish Sastry, Amanda Askell,
  Sandhini Agarwal, Ariel Herbert-Voss, Gretchen Krueger, Tom Henighan, Rewon
  Child, Aditya Ramesh, Daniel Ziegler, Jeffrey Wu, Clemens Winter, Chris
  Hesse, Mark Chen, Eric Sigler, Mateusz Litwin, Scott Gray, Benjamin Chess,
  Jack Clark, Christopher Berner, Sam McCandlish, Alec Radford, Ilya Sutskever,
  and Dario Amodei.
\newblock Language models are few-shot learners.
\newblock In H.~Larochelle, M.~Ranzato, R.~Hadsell, M.F. Balcan, and H.~Lin,
  editors, {\em Advances in Neural Information Processing Systems}, volume~33,
  pages 1877--1901. Curran Associates, Inc., 2020.

\bibitem{physPrev1}
Gerd Kortemeyer.
\newblock Could an artificial-intelligence agent pass an introductory physics
  course?
\newblock {\em Phys. Rev. Phys. Educ. Res.}, 19:010132, May 2023.

\bibitem{physPrev2}
Kay Lehnert.
\newblock Ai insights into theoretical physics and the swampland program: A
  journey through the cosmos with chatgpt.
\newblock {\em arXiv preprint arXiv:2301.08155}, 2023.

\bibitem{mathsCapChat1}
Simon Frieder, Luca Pinchetti, Ryan-Rhys Griffiths, Tommaso Salvatori, Thomas
  Lukasiewicz, Philipp~Christian Petersen, Alexis Chevalier, and Julius Berner.
\newblock Mathematical capabilities of chatgpt.
\newblock {\em arXiv preprint arXiv:2301.13867}, 2023.

\bibitem{mathsCapChat2}
Paulo Shakarian, Abhinav Koyyalamudi, Noel Ngu, and Lakshmivihari Mareedu.
\newblock An independent evaluation of chatgpt on mathematical word problems
  (mwp).
\newblock {\em arXiv preprint arXiv:2302.13814}, 2023.

\bibitem{jinja}
Jinja2, April 2022.
\newblock \url{https://pypi.org/project/Jinja2/}.

\bibitem{foi_2018}
James Andrew.
\newblock Undergraduate module averages 2018 - a freedom of information request
  to university of durham, November 2018.

\bibitem{foi_2019}
Jon Ruislip.
\newblock Undergraduate module averages 2019 - a freedom of information request
  to university of durham, August 2019.

\bibitem{foi_2020}
T~Hall.
\newblock Undergraduate module averages 2020 - a freedom of information request
  to university of durham, July 2021.

\bibitem{foi_2021}
Alex Ross.
\newblock Durham university module averages 2021 - a freedom of information
  request to university of durham, November 2021.

\bibitem{humanDetection1}
Maurice Jakesch, Jeffrey~T Hancock, and Mor Naaman.
\newblock Human heuristics for ai-generated language are flawed.
\newblock {\em Proceedings of the National Academy of Sciences},
  120(11):e2208839120, 2023.

\bibitem{humanDetection2}
Mike Perkins.
\newblock Academic integrity considerations of ai large language models in the
  post-pandemic era: Chatgpt and beyond.
\newblock {\em Journal of University Teaching \& Learning Practice}, 20(2):07,
  2023.

\bibitem{humanDetection3}
Nils K{\"o}bis and Luca~D Mossink.
\newblock Artificial intelligence versus maya angelou: Experimental evidence
  that people cannot differentiate ai-generated from human-written poetry.
\newblock {\em Computers in human behavior}, 114:106553, 2021.

\bibitem{detectAI}
Vinu~Sankar Sadasivan, Aounon Kumar, Sriram Balasubramanian, Wenxiao Wang, and
  Soheil Feizi.
\newblock Can ai-generated text be reliably detected?
\newblock {\em arXiv preprint arXiv:2303.11156}, 2023.

\bibitem{englishBias}
Weixin Liang, Mert Yuksekgonul, Yining Mao, Eric Wu, and James Zou.
\newblock Gpt detectors are biased against non-native english writers.
\newblock {\em arXiv preprint arXiv:2304.02819}, 2023.

\end{thebibliography}

\end{document}